\documentclass[preprint,10pt]{elsarticle}

\usepackage{amssymb}
\usepackage{bm}

\journal{Physics Letters A}

\begin{document}

\begin{frontmatter}



\title{Superconductivity from the Berry connection from many-electron wave functions: $U(1)$ gauge fluctuation, gauge invariance, and superconductivity}


\author{Hiroyasu Koizumi}

\affiliation{organization={Center for Computational Sciences, University of Tsukuba},
            addressline={1-1-1 Tennodai}, 
            city={Tsukuba},
            postcode={305-8577}, 
            state={Ibaraki},
            country={Japan}}

\begin{abstract}
The electromagnetic $U(1)$ gauge invariance and superconductivity are examined by employing the Berry connection from many-electron wave functions
and the fluctuations of it that give rise to the probabilistic random transitions among quantum states.
The gauge invariance in the classical electrodynamics is regarded as the requirement that extracts
unaffected parts by the gauge fluctuation. In quantum mechanics, dissipation-free states from the gauge fluctuation exist,
and we identify them as superconducting states. We explain the reversible superconducting-normal phase transition in a magnetic field 
including the force balance.
The present superconductivity  theory lifts two fundamental disagreements between theory and experiment existing in the current standard theory with keeping the major results by the BCS theory. Thus, it may be considered as a superconductivity theory encompassing the current standard one.
\end{abstract}







\end{frontmatter}



\section{Introduction}

After the discovery of the electron \cite{JJThomson}, a free electron model of electric conduction was put forward by Drude \cite{Drude}, and further elaborated by Lorentz \cite{Lorentz}.
In this model, current carrying electrons in a metal are treated as free particles with charge $-e$ and mass $m$ under the influence of an electric field ${\bf E}$ and a friction force linearly depends on the velocity ${\bf v}$
\begin{eqnarray}
m{d \over {dt}}{\bf v}=-{m \over \tau}{\bf v}-e{\bf E}
\label{Drude}
\end{eqnarray}
where $\tau$ is the relaxation time by the friction. It is assumed that collisions give rise to $\tau$.
The Lorentz extended it by including 
the force from the magnetic field $-e {\bf v} \times {\bf B}$.

The normal state of the current standard superconductivity theory is the quantum version of the Drude model;
electron-electron interactions are added to it, and electron-pairs are generated from the conduction electrons to give rise to the energy gap.
This electron-pairing energy gap is the hallmark of the superconducting state in the standard theory \cite{BCS1957},
and theories have been developed for accurate calculations of it.

The expression in Eq.~(\ref{Drude}) is viewed, differently, in the fluctuation-dissipation theory. The dissipation term $-{m \over \tau}{\bf v}$ arises from 
fluctuating forces that exist in the system. 
Taking into account the electric field fluctuation, Kubo obtained the electric conductivity \cite{Kubo1957}.
In this case, the parameter $\tau$ contains the many-body effect.
In the following, we adopt this view, and take
the electric current as a quantum fluid with the velocity field ${\bf v}$ under the influence of background fluctuating forces.

According to the Ginzburg-Landau theory \cite{GL} the velocity field ${\bf v}$ in the superconductor is expressed as
\begin{eqnarray}
{\bf v}=-{Q \over M} \left({\bf A} - {\hbar \over {2e}} \nabla \chi \right)
\label{velocity-field}
\end{eqnarray}
where $Q$ is the supercurrent carrier charge and $M$ is its mass; $\chi$ is an angular variable
with period $2\pi$, which is identified as the phase of the superconducting order parameter. The factor ${\hbar \over {2e}}$ in front of $\nabla \chi$ is so chosen that the observed flux quantization is obtainable.
Note that the vector potential ${\bf A}$ appears explicitly in ${\bf v}$, although it
is considered as a computational mathematical tool to express the electric field ${\bf E}$ and magnetic filed ${\bf B}$
in classical dynamics. It is now confirmed that it is a physical entity in quantum mechanics \cite{AB1959,Tonomura1986}.
Note also that $\chi$ corresponds to the Nambu-Goldstone mode that saves the gauge invariance of the supercurrent, which was discovered by Nambu  \cite{Nambu1960} using the Bogoliubov formalism \cite{Bogoliubov58} of the BCS theory \cite{BCS1957}.

The above ${\bf v}$ yields the London moment magnetic field (the magnetic field generated inside a rotating superconductor)
given by
\begin{eqnarray}
{\bf B}_{\rm London}=-{{2M} \over Q} {\bm \omega}
\end{eqnarray}
where  ${\bm \omega}$ is the angular velocity of the rotation \cite{London1950,koizumi2021}.
If we follow the standard superconductivity theory, we have $Q=-2e$ and $M=2m^\ast$
assuming supercurrent carriers are electron-pairs of the effective mass $m^\ast$.
Then, 
${\bf B}_{\rm London}={{2m^\ast} \over e} {\bm \omega}$ is obtained; however, this actually disagrees with the experimental result, which yields the free electron mass $m_e$ instead of $m^\ast$ \cite{Tate1989,Tate1990,VERHEIJEN1990a,Verheijen1990,Sanzari1996}.
A new theory that explains $m_e$ (not $m^\ast$) in ${\bf B}_{\rm London}$ has been put forward by the present author \cite{koizumi2021}. It attributes $\chi$ to the Berry phase from many-electron wave functions. 

Let us consider the velocity field in the new theory. We take the total derivative of ${\bf v}$ in Eq.~(\ref{velocity-field}), which yields the following equation similar equation to Eq.~(\ref{Drude})
\begin{eqnarray}
{{d{\bf v}} \over {dt}}&=&\partial_t{\bf v}+({\bf v} \cdot \nabla){\bf v}
=\partial_t{\bf v}+
{1 \over 2}\nabla v^2-{\bf v}\times(\nabla \times {\bf v})
\nonumber
\\
&=&-{ Q \over {M}} 
\left[\partial_t {\bf A}+{ \hbar \over {2e}}\partial_t(\nabla \chi)
-{\bf v}\times{\bf B}
+{ \hbar \over {2e}}{\bf v}\times \nabla \times (\nabla \chi) \right]+{1 \over 2}\nabla v^2
\nonumber
\\
&=&{ Q \over {M}} 
\left[{\bf E}+{\bf v}\times{\bf B}-{ \hbar \over {2e}}\partial_t(\nabla \chi)
+{ \hbar \over {2e}}(\nabla \partial_t \chi)
+{ \hbar \over {2e}}{\bf v}\times \nabla \times (\nabla \chi) \right]-{1 \over M} \nabla \mu+{1 \over 2}\nabla v^2
\nonumber
\\
\label{eq13a}
\end{eqnarray}
where ${\bf E}=-\partial_t {\bf A} -\nabla \varphi$
and ${\bf B}=\nabla \times {\bf A}$ are employed \cite{Koizumi2024Lorentz};
$\mu$ is the chemical potential given by
\begin{eqnarray}
\mu=-Q \left( \varphi +{ \hbar \over {2e}}\partial_t \chi \right)
\label{eq-chemical}
\end{eqnarray}
There are extra terms in addition to the standard Lorentz force  \cite{Koizumi2024Lorentz}.
They are shown to be important to explain supercurrent related phenomena.

 The main concern of the present work is `$\partial_t(\nabla \chi)$' seen in Eq.~(\ref{eq13a}).
We call it the {\em gauge fluctuation}. 

It is shown that $\nabla \chi$ is obtained from the following two constraints \cite{Koizumi2024Lorentz}, 
\begin{enumerate}
\item The wave function is a single-valued function of the electron coordinates.

\item The local charge is conserved.

\label{item3}
\end{enumerate}

When probabilistic random transitions among quantum states occur, the fluctuation of $\nabla \chi$ should occur 
due to the requirement 2 of the given list
since the local charge distribution is affected by the change of states caused by such transitions.
  In this work, we will take into account the fluctuation of $\nabla \chi$ by treating $\chi$ as a random field. 

In order to consider the quantum effect from $\chi$, we consider the following fact: We may view ${\bf A}$ in Eq.~(\ref{velocity-field}) as a gauge-transformed one from ${\bf A}_0$
\begin{eqnarray}
{\bf A}= {\bf A}_0+{\hbar \over {2e}}\nabla \chi, \quad  \phi=\phi_0 - {\hbar \over {2e}}{{\partial \chi} \over {\partial t}} 
\end{eqnarray}
When the fluctuations of  $\nabla \chi$ and $\partial_t \chi$ exist, ${\bf A}$ and $\phi$ are not stable quantities.
On the other hand ${\bf B}$ and ${\bf E}$ are stable. Thus, we may say that the gauge invariance requirement
is the procedure to extract the unaffected parts from the gauge fluctuation.

In the present work, the quantum version of the gauge fluctuation is investigated using the Lindblad equation \cite{VanKampen,QuantumNoise}.
We show that there are quantum states that are free from the dissipation by the gauge fluctuation.
We identify them as superconducting states.
We also explain the non-dissipative superconducting-normal phase transition in a magnetic field using the phase variable $\chi$\cite{koizumi2020b,koizumi2024}. These results indicate that we may take the stable $\nabla \chi$  as the hallmark of superconducting states.

\section{Berry connection from many-electron wave functions}

We start with the following Langevin equation \cite{VanKampen,QuantumNoise} to describe the dissipative electric current flow in a metallic wire
\begin{eqnarray}
m_e { {d{\bf v}} \over {dt}}=-{m_e \over \tau} {\bf v}+{\bm \eta}
\label{eqLangevin}
\end{eqnarray}
Kubo considered an electric field as a noise ${\bm \eta}$; however, we consider the case ${\bm \eta}=(\eta_i, \eta_2, \eta_3)={\hbar \over 2}\partial_t \nabla \chi$. The thermal average of it satisfies the following relations
\begin{eqnarray}
\langle \eta_i(t) \rangle=0, \quad \langle \eta_i(t) \eta_j(t') \rangle=2\epsilon \delta_{i j } \delta(t-t')
\label{Langevin2}
\end{eqnarray}
where $\epsilon$ is the strength of the fluctuation.
The relaxation time $\tau$ is obtained as $\tau={\epsilon ({k_B T})^{-1}}$ when the classical equipartition result is used. 
This indicates that the current generated by the velocity field in Eq.~(\ref{velocity-field}) is zero if the fluctuation of $\nabla \chi$ exists.

The fact that supercurrent with the velocity field in Eq.~(\ref{velocity-field}) exists in a superconducting state indicates that the superconducting state is a state with a stable $\nabla \chi$. 
We attribute this $\chi$ to the Berry connection arising from many-electron wave functions 
as will be explained in the following.

The Lagrangian with the vector potential ${\bf A}_0$ is given by
\begin{eqnarray}
L_0=\sum_{j=1}^N \left[ {m_e \over {2}}\dot{\bf r}_j^2-\dot{\bf r}_j \cdot {\bf A}_0
+e\phi_0({\bf r}_j,t)-V_1({\bf r}_j) \right]
-\sum_{j>k}V_2({\bf r}_j, {\bf r}_k)
\end{eqnarray}
where $V_1$ is the single-particle potential, and $V_2$ the two-particle one; and that with ${\bf A}$
is given by
\begin{eqnarray}
L&=&L_0 -e \sum_{j=1}^N {\hbar \over {2e}} \left[ \partial_t \chi({\bf r}_j,t)+ \dot{\bf r}_j \cdot \nabla \chi({\bf r}_j,t) \right]
=L_0 -{d \over {dt}} \sum_{j=1}^N {\hbar \over {2}}\chi({\bf r}_j,t)
\end{eqnarray}
Since the least action principle yields the same classical equations-of-motion even
a total time-derivative term is added to the Lagrangian,
$L$ and $L_0$ describe the same classical dynamics; thus, the gauge invariance occurs in classical mechanics.

Now we shall consider a quantum mechanical version. We construct the Hamiltonian; the one obtained from $L_0$ is given by
\begin{eqnarray}
H_0&=&\sum_{j=1}^N \left\{{1 \over {2m_e}} \left[{\bf p}_j+e{\bf A}_0\chi({\bf r}_j,t) \right]^2
-e\phi_0({\bf r}_j,t)
+V_1({\bf r}_j) \right\} + \sum_{j>k}V_2({\bf r}_j, {\bf r}_k)
\nonumber
\\
\end{eqnarray}
and that from $L$ is 
\begin{eqnarray}
H&=&\sum_{j=1}^N \left\{{1 \over {2m_e}}
\left[{\bf p}_j+e{\bf A}_0\chi({\bf r}_j,t)+{\hbar \over {2}}\nabla \chi({\bf r}_j,t) \right]^2
-e\phi_0({\bf r}_j,t)+{\hbar \over {2}}\partial_t\chi({\bf r}_j,t) 
+V_1({\bf r}_j) \right\}
\nonumber
\\
&+&\sum_{j>k}V_2({\bf r}_j, {\bf r}_k)
\end{eqnarray}

 Let us define the following unitary matrix $U$
\begin{eqnarray}
U=e^{-{i \over 2}\sum_j \chi({\bf r}_j,t)}
\label{transform}
\end{eqnarray}
Then, the wave function before the gauge transformation $\Psi_0$ and that after it $\Psi$ are related as follows
\begin{eqnarray}
\Psi=U\Psi_0
\label{eqUnitary}
\end{eqnarray}
with the following change in the Scr\"{o}dinger equation 
\begin{eqnarray}
i \hbar \partial_t \Psi_0 = H_0 \Psi_0 \ \rightarrow \
i \hbar \partial_t \Psi= H \Psi
\end{eqnarray}
Since the transformation by $U$ in Eq.~(\ref{transform}) is unitary, two Scr\"{o}dinger equations
describe the same physics.
However, for the case where $U$ appears as a Berry phase, and a quantization of the field arising from it 
is performed, the transformation becomes  non-unitary, as will be explained below.

Let us take the total wave function $\Psi({\bf x}_1, \cdots, {\bf x}_N)$, where ${\bf x}_i$ collectively stands for the coordinate ${\bf r}_i$ and the spin $\sigma_i$ of the $i$th electron; $N$ is the total number of electrons.
Now we consider the Berry phase arising from the constrains that ``$\Psi({\bf x}_1, \cdots, {\bf x}_N)$ is single-valued with respect to ${\bf r}_1, \cdots, {\bf r}_N$ and the current calculated from it
satisfies the conservation of the local charge'' (the two constraints listed in Introduction). 

The essential ingredient of the new theory is the 
Berry connection from many-body wave functions defined by
\begin{eqnarray}
\!{\bf A}^{\rm MB}_{\Psi}({\bf r},t)\!=\!
{{{\rm Re} \left\{
 \int d\sigma_1  d{\bf x}_{2}  \cdots d{\bf x}_{N}
 \Psi^{\ast}({\bf r}, \sigma_1, \cdots, {\bf x}_{N},t)
  (-i \hbar \nabla )
\Psi({\bf r}, \sigma_1, \cdots, {\bf x}_{N},t) \right\}
 }
 \over {\hbar N^{-1} \rho({\bf r},t)}} 
 \label{eq16}
\end{eqnarray}
taking ${\bf r}$ as the `adibatic' parameter in the usual Berry connection \cite{Berry,BMKNZ}, although we do not consider adiabticity. It gives rise to a many-body effect which is missing in a current standard many-body theory.
Here,
`$\rm{Re}$' denotes the real part,   $-i \hbar \nabla$ is the Schr\"{o}dinger's momentum operator for the coordinate vector ${\bf r}$. and $\rho$ is the number density calculated from $\Psi$.

We identify $\chi$ as follows
  \begin{eqnarray}
{ {\chi({\bf r},t)}}= - 2\int^{{\bf r}}_0 {\bf A}_{\Psi}^{\rm MB}({\bf r}',t) \cdot d{\bf r}' 
\label{eq17}
\end{eqnarray}
The many-electron wave function $\Psi$ is expressed as
\begin{eqnarray}
\Psi({\bf x}_1, \cdots, {\bf x}_N,t)=\exp \left(  -{i \over 2} \sum_{j=1}^{N} \chi({\bf r}_j,t)\right)
\Psi_0({\bf x}_1, \cdots, {\bf x}_N, t)
\label{single-valued}
\end{eqnarray}
with $\Psi_0$ being a currentless wave function \cite{koizumi2019,koizumi2022}.

\section{Quantization of $\chi$}

The variable $\chi$ arises to satisfy the constraints; we may treat it as an independent variable as in the case for constrained Hamiltonian dynamics \cite{Dirac1958,Koizumi2024Lorentz}.
We use the following Lagrangian to obtain the conjugate field of $\chi$
\begin{eqnarray}
L=\langle \Psi | i\hbar {\partial \over {\partial t}} -H |\Psi \rangle
=\int d^3 r { \hbar \over 2} \dot{\chi}  \rho + i\hbar \langle \Psi_0 |{\partial \over {\partial t}}|\Psi_0 \rangle-\langle \Psi | H |\Psi \rangle
\label{eq19}
\end{eqnarray}
The conjugate field to $\chi$, $\pi_{\chi}$, is obtained as
\begin{eqnarray}
\pi_{\chi}={{\delta L} \over {\delta \dot{\chi} }}={ \hbar \over 2} \rho
\end{eqnarray}
The canonical quantization condition $[\hat{\chi}({\bf r},t),\hat{\pi}_{\chi} ({\bf r}',t)]=i\hbar \delta^{(3)}({\bf r}-{\bf r}')$
yields
\begin{eqnarray}
[\hat{\chi}({\bf r},t),\hat{\rho}({\bf r}',t)]=2i\delta^{(3)}({\bf r}-{\bf r}')
\label{eq16-4-68}
\end{eqnarray}
where $\hat{\chi}$ and $\hat{\rho}$ are operators corresponding to ${\chi}$ and ${\rho}$, respectively. 

In the wave function in Eq.~(\ref{single-valued}),
all electrons participate in the collective mode describe by $e^{-{i \over 2}\chi}$ (we call it `{\em $\chi$ mode}').
However, the total energy can be reduced by
allowing the fluctuation of the numbers participating in the $\chi$ mode.
In this case, $\Psi$ is expressed as
\begin{eqnarray}
\Psi({\bf x}_1, \cdots, {\bf x}_{N},t)&=&C_0 \Psi_{00}({\bf x}_1,  \cdots, {\bf x}_{N},t)
\nonumber
\\
&+&
C_1 \sum_{j_1=1}^{N} e^{  -{i \over 2} \chi({\bf r}_{j_1},t)}\Psi_{01}({\bf x}_1,  \cdots, {\bf x}_{N},t)
\nonumber
\\
&+&C_2 \sum_{j_1 > j_2} e^{-{i \over 2} [\chi({\bf r}_{j_1},t) + \chi({\bf r}_{j_2},t) ]}\Psi_{02}({\bf x}_1,  \cdots, {\bf x}_{N},t)
\nonumber
\\
&+&C_3 \sum_{j_1 > j_2 > j_3} e^{  -{i \over 2} [\chi({\bf r}_{j_1},t) + \chi({\bf r}_{j_2},t) 
+ \chi({\bf r}_{j_3},t)]}\Psi_{03}({\bf x}_1,  \cdots, {\bf x}_{N},t)
\nonumber
\\
&&\cdots \cdots
\nonumber
\\
&+&C_{N} e^{ -{i \over 2} [ \chi({\bf r}_1,t)+ \cdots + \chi({\bf r}_N,t) ]}
\Psi_{0{N}}({\bf x}_1,  \cdots, {\bf x}_{N},t)
\label{eq16-4-31}
\end{eqnarray}
We replace the requirement  from `the wave function is a single-valued function of the all electron coordinates'
to `the wave function is a single-valued function of current carrying electron coordinates', and 
employ above $\Psi$ in Eq.~(\ref{eq19}). Then, the electron density participating in the $\chi$ mode is reduced from $\rho$ (we denote the reduced one by  $n_s$).
Then, the commutation relation in Eq.~(\ref{eq16-4-68}) becomes 
\begin{eqnarray}
[\hat{\chi}({\bf r},t),\hat{n_s}({\bf r}',t)]=2i\delta^{(3)}({\bf r}-{\bf r}')
\end{eqnarray}
From this relation, the number changing operators
$e^{{ i \over 2} \hat{\chi}({\bf r},t)}$ that increases the number of electrons in the $\chi$-mode by one,
and $e^{-{ i \over 2} \hat{\chi}({\bf r},t)}$ that decreases by one are constructed.
A practical method to include this operator for the calculation of superconducting states
was explained in our previous work \cite{Koizumi2021c}. It uses the particle-number conserving Bogoliubov-de Gennes equations,
which reproduces the major results obtained by the standard theory.

Now let us consider the dissipative process involving $\hat{\chi}$.
For the purpose, we consider the following quantum mechanical version of the Langevin equation ({\em Schr\"{o}dinger-Langevin Equation})
\begin{eqnarray}
{d \over {dt}}|{\Psi}(t)\rangle=-i \hbar^{-1}H |\Psi(t)\rangle -\epsilon V^\dagger V  |\Psi(t)\rangle + \eta(t)V|\Psi(t)\rangle
\label{eq17-7-1}
\end{eqnarray}
which includes $\eta(t)V$ as a stochastic potential \cite{VanKampen};
$\eta(t)$ describes the quantum mechanical probabilistic jumps that satisfies
\begin{eqnarray}
\langle \eta(t) \rangle=0, \quad \langle \eta(t) \eta^\ast(t') \rangle=2\epsilon \delta(t-t')
\label{eq17-7-3}
\end{eqnarray}
where $\epsilon$ is the frequency of the jumps.
From the above, the following equation for the density matrix $\rho_d(t)=|{\Psi}(t)\rangle \langle {\Psi}(t)|$ is obtained 
\begin{eqnarray}
{{d \rho_d(t)} \over {dt}}=-i \hbar^{-1}  [H, \rho_d(t)] 
+{\epsilon} \left[ 2V \rho_d(t)V^\dagger -  \rho_d(t)
V^\dagger V
- V^\dagger V \rho_d(t)
\right]
\end{eqnarray}
This is a Lindblad equation.

We extend it to the case where more than one $V$ denoted as $V_\alpha$ exists.
Then, we have
\begin{eqnarray}
{{d \rho_d(t)} \over {dt}}=-i \hbar^{-1}  [H, \rho_d(t)] 
+\sum_\alpha {\epsilon_\alpha} \left[ 2V_\alpha \rho_d(t)V_\alpha^\dagger -  \rho_d(t)
V_\alpha^\dagger V_\alpha 
- V_\alpha^\dagger V_\alpha  \rho_d(t)
\right]
\label{Lindblad}
\end{eqnarray}
where $V_\alpha$ is the quantized contribution from Eq.~(\ref{transform}) for the $\alpha$th electron 
\begin{eqnarray}
V_{\alpha}=e^{-{i \over 2} \hat{\chi}({\bf r}_\alpha)}
\end{eqnarray}
When transitions between electronic states occur, the conservation condition for the local charge changes, resulting in the modification of $\chi$. 
We will take into account such changes as random probabilistic changes described by $V_{\alpha}$.

Let us examine the solution of Eq.~(\ref{Lindblad}). Non-dissipative solutions occur in two cases.
One is where  $|{\Psi}(t)\rangle $ does not contain electrons participating in the $\chi$-mode;
in this case, $L_\alpha \rho_d(t)$ and $\rho_d(t)L_\alpha^\dagger$ are both zero.
Another case occurs when $|{\Psi}(t)\rangle $ is the eigenstate of $V_\alpha$, i.e. it satisfies
\begin{eqnarray}
e^{-{i \over 2} \hat{\chi}({\bf r}_\alpha)}|\chi \rangle =e^{-{i \over 2} {\chi}({\bf r}_\alpha)}|\chi \rangle
\end{eqnarray}
where the eigenket $|{\Psi}(t)\rangle $ in this case is denoted as $|\chi \rangle$.
We identify such a state as a superconducting state. It gives rise to the supercurrent with the velocity field given in Eq.~(\ref{velocity-field}).

\section{Finite temperature case}
\label{Sec4}
We now consider a general case where the Berry connection arises from a set of states $\{ \Psi_j \}$
with a probability distribution $p_j$ for $\Psi_j$ that satisfy
\begin{eqnarray}
\sum_j p_j=1
\end{eqnarray}
In this case, we may consider the average Berry connection given by
\begin{eqnarray}
{\bf A}^{\rm MB}=\sum_j p_j {\bf A}_{\Psi_j}^{\rm MB}
\end{eqnarray}
where ${\bf A}_{\Psi_j}^{\rm MB}$ is obtained from Eq.~(\ref{eq16}) by replacing $\Psi$ with $\Psi_j$.
Then, we may use this ${\bf A}^{\rm MB}$ for ${\bf A}_\Psi^{\rm MB}$ in Eq.~(\ref{eq17}). The finite temperature case corresponds to the case where $p_j$ is given by the Boltzmann distribution.

\section{Non-dissipative superconducting-normal phase transition in a magnetic field}

We now consider the superconducting-normal state transition in a magnetic field using 
a part of the Ginzburg-Landau free energy given as a sum of the magnetic field energy and the kinetic energy \cite{GL,koizumi2020b}.
Note that there is serious confusion and inconsistency in the Free energy expression \cite{Nikulov2025,Hirsch2026}.
In the following, we employ the Ginzburg-Landau version that includes both material and magnetic field contributions.
In this free energy, the conversion occurs between the electron kinetic energy in the surface region and that of the magnetic field
upon the first order superconducting-normal phase transition in a magnetic field. On the other hand, the Gibbs free energies of the two phases at the transition point are equal. 

The magnetic field energy is given by
\begin{eqnarray}
F_m={ 1 \over {2 \mu_0}} \int d^3 r \ {\bf B}^2
\end{eqnarray}
where $\mu_0$ is the vacuum permeability,
and the kinetic energy is given by
\begin{eqnarray}
F_k={ M \over {2}} \int d^3 r \ {\bf v}^2 n_s 
\end{eqnarray}

The change of the magnetic field energy during a time interval $0 < t <\Delta t$ is given by
\begin{eqnarray}
\Delta F_m={ 1 \over  {\mu_0}}\int_0^{\Delta t} dt  \int d^3 r \ {\bf B}\cdot{ {\partial {\bf B}} \over {\partial t}}
=
-\int_0^{\Delta t} dt  \int d^3 r \ {\bf J}\cdot {\bf E}
\label{delta1}
\end{eqnarray}
where  ${\bf J}_s=Q{\bf v}n_s$ is the current density, and two of Maxwell's equations
$\nabla \times {\bf E}=-{ {\partial {\bf B}} \over {\partial t}}$
and 
 $ \nabla \times {\bf B}=\mu_0 {\bf J}$
are used for the derivation. 
$\Delta F_m$ contains the term ${\bf J}\cdot {\bf E}$, thus, it will produce the Joule heat. 

In the standard theory of superconductivity it is assumed that the
electron-pairs flow without dissipation, but single electrons flow with dissipation. Then, near the transition point of the superconducting-normal phase transition, a significant number of electron-pairs are broken into single electrons, leading to the production of the Joule heat.
This however is against the fact that the superconducting-normal phase transition in a magnetic field is a reversible process for the type-I superconductors. It is also important to explain the force balance between the transition.
 We will show that this Joule heating is avoided in the present theory with maintaining the force balance, below.

Let us consider the change of the kinetic energy due to the change of ${\bf v}$ during a time interval $0 < t <\Delta t$; it  is given by
\begin{eqnarray}
\Delta F_k&=&M\int_0^{\Delta t} dt   \int d^3 r \ n_s {\bf v} \cdot  { {\partial {\bf v}} \over {\partial t} }
\nonumber
\\
&=&-\int_0^{\Delta t} dt   \int d^3 r \  {\bf J} \cdot \partial_t \left( {\bf A} -{\hbar \over {2e}} \nabla \chi \right)
\label{delta2}
\end{eqnarray}

The reversible process implies $\Delta F_m+\Delta F_k=0$. Using Eqs.~(\ref{delta1}) and (\ref{delta2}), 
the non-dissipative transition occurs when the following condition is satisfied
\begin{eqnarray}
0={\bf E}+\partial_t \left( {\bf A} -{\hbar \over {2e}} \nabla \chi \right)=Q^{-1} \nabla \mu
\end{eqnarray}
where $\mu$ is given in Eq.~(\ref{eq-chemical}).
The above indicates that the energy balance is satisfied due to the fact that the Gibbs free energies of the two phases at the transition point are equal. It also enables the transition with $\Delta F_m+\Delta F_k=0$.
This non-dissipative transition becoems possible due to fact that the supercurrent is a flow generated by
the velocity field ${\bf v}$ in Eq.~(\ref{velocity-field}), not the current generated by the electron-pair motion
in a strict sense although it appears so. 
It is noteworthy that the so-called `quasiparticle poisoning problem' indicates the existence a large amount of excited single electrons in Josephson junction systems, obtaining the observed ratio of their number to the Cooper pair number  $10^{-9} \sim 10^{-5}$ in disagreement with the standard theory predicted value $10^{-52}$\cite{poisoning2023,Serniak2019}. The supercurrent should be considered as a current generated by
the velocity field that includes stable $\nabla \chi$.

\section{Concluding remarks}
\label{Sec5}

Previously, it was shown that the present theory explains the major results obtained by the BCS theory  \cite{koizumi2022,Koizumi_2023}.
The present work indicates that  it also explains the two problems that disagree with the current standard theory.
Thus, we may say that the new theory is a theory that encompasses the BCS one.

The current standard microscopic theory of superconductivity was originated from
the BCS theory \cite{BCS1957}. The hallmark of the superconducting state in this theory
is the electron-pair formation. 
The Meissner effect was explained by the rigidity of the wave function arising from the energy gap
created by the electron-pair formation; and the presence of the supercurrent was claimed to be included in the
explanation of the Meissner effect.

However, the gauge invariance problem remained in the original derivation of the Meissner effect in the BCS theory.
To restore the gauge invariance, the mode arising from the fluctuation of the number of electron-pairs was required \cite{Anderson1958a,Anderson1958b,Nambu1960}.
Actually, Anderson and Weinberg \cite{Anderson66,Weinberg} stressed that `fluctuation of the number of the supercurrent carriers
and the appearance of its conjugate phase variable' 
is the most important characteristics of superconducting states. 
The present theory follows their viewpoint. However, the way the phase variable appears 
is different.
The fact that the phase variable  was not correctly included in the standard theory  is manifested in the two problems that contract experimental results.
In the present work, we identify it as the one that originates from the Berry connection arising from many-electron wave functions.
In this theory, the stable $\nabla \chi$ is the hallmark of the superconducting state.
We would like to point out that the experimental superconducting transition temperature for the high transition temperature cuprate superconductors does not corresponds to the energy gap formation temperature but to
the stabilization temperature of nano-sized loop currents \cite{Kivelson95}, and a theoretical simulation exits
that indicates it corresponds to the stabilization temperature of $\nabla \chi$ \cite{HKoizumi2015B}.

 \bibliographystyle{elsarticle-num} 

\bibliography{SPIN-BCS}





\end{document}